\documentclass[twoside,leqno,twocolumn]{article}

\usepackage[letterpaper]{geometry}
\usepackage{amssymb}
\usepackage{amsmath}
\usepackage{ltexpprt}
\usepackage{hyperref}
\usepackage{graphicx}
\usepackage{booktabs}
\usepackage{multirow}
\usepackage{caption}
\usepackage{subcaption}
\usepackage{float}
\usepackage{fancyhdr}
\usepackage{balance}
\usepackage{comment}
\usepackage{xcolor}
\usepackage{xspace}
\usepackage{enumitem}
\usepackage{algorithmicx}
\usepackage{algorithm}
\usepackage{algpseudocode}
\algtext*{EndWhile}
\algtext*{EndIf}
\algtext*{EndFunction}
\algtext*{EndFor}
\algrenewcommand\algorithmicrequire{\textbf{Input:}}
\algrenewcommand\algorithmicensure{\textbf{Output:}}

\newcommand*{\stat}{population statistics }
\newcommand*{\geo}{geo }

\newcommand\nima[1]{\textcolor{red}{(nima) #1}\xspace}

\newcommand{\system}{{\sc PopSim}\xspace}

\begin{document}

\newcommand\relatedversion{}


\title{\system: An Individual-level Population Simulator for \\Equitable Allocation of City Resources\thanks{This project was supported in part by NSF 2107290.}}


\author{Khanh Duy Nguyen, Nima Shahbazi, Abolfazl Asudeh\thanks{University of Illinois Chicago \newline
\{knguye71, nshahb3, asudeh\}@uic.edu}}

\date{}

\maketitle

\fancyfoot[R]{\scriptsize{Copyright \textcopyright 2023 by SIAM
Unauthorized reproduction of this article is prohibited}}





\begin{abstract} \small\baselineskip=9pt 
Historical systematic exclusionary tactics based on race have forced people of certain demographic groups to congregate in specific urban areas. Aside from the ethical aspects of such segregation, these policies have implications for the allocation of urban resources including public transportation, healthcare, and education within the cities. The initial step towards addressing these issues involves conducting an audit to assess the status of equitable resource allocation. However, due to privacy and confidentiality concerns, individual-level data containing demographic information cannot be made publicly available. By leveraging publicly available aggregated demographic statistics data, we introduce \system, a system for generating semi-synthetic individual-level population data with demographic information. We use \system to generate multiple benchmark datasets for the city of Chicago and conduct extensive statistical evaluations to validate those. We further use our datasets for several case studies that showcase the application of our system for auditing equitable allocation of city resources.
\end{abstract}
\section{Introduction}
Over the past several decades, the fast pace of urbanization has caused a sharp rise in city resource consumption. This not only affects residents’ welfare levels but also plays a crucial role in shaping the sustainability of urban services and development.
Unfortunately, a range of political and cultural forces motivated by racist attitudes towards people of color have created lines of separation between the citizens, effectively segregating people from different racial and ethnic groups into specific areas of the city.  
The impact of segregation remains particularly strong in cities like Chicago, which is home to some of the most deeply segregated areas that were once designated as ``redlined'' areas~\cite{nardone2020historic}. Such systematic exclusionary tactics have also been reflected in urban planning and resource allocation policies creating inequitable access levels to services such as public transport, healthcare, and education among different demographic groups. 

Several organizations and research groups from a variety of disciplines, such as {STEM (Science, Technology, Engineering, and Mathematics)}, social sciences, economics, and urban planning, are actively engaged in efforts to counteract the impact of the aforementioned tactics. {More specifically, socially fair, just, and equitable resource allocation has gained significant attention in light of the growing concern for fairness in computational problems~\cite{blanco2022fairness,asudeh2022maximizing,lan2010axiomatic}.} However, prior to any interventions, it is crucial to conduct a comprehensive audit of the current allocation of resources within the urban environment.
Auditing may require access to individual-level data with demographic information. Unfortunately, obtaining such data is often challenging, as it is not publicly available due to concerns regarding privacy and confidentiality. Luckily, there is information readily available on aggregated demographic statistics for various neighborhoods (down to the block level in a city), providing a general overview of demographic distributions. This motivates us to build \system, a system for creating simulated individual-level data that is consistent with the available demographic statistics and can be utilized for auditing resource allocation within the city. 
In the development of our system, we use techniques such as Inverse-CDF~\cite{devroye1986sample} and Monte Carlo rejection sampling~\cite{raeside1976monte} to draw unbiased samples from different demographic groups and granularity levels in the geo-location hierarchies, ranging from an entire city down to a specific coordinate.
\system enables its users to not only sample individuals but also generate large-size individual-level semi-synthetic datasets. It also enables designing a wide range of randomized and sampling-based algorithms for equitable allocation of resources.

To showcase an application of \system, we fine-tune it for Chicago and generate several semi-synthetic datasets with various sizes.
We use statistical tests to validate that the simulated data follows the publicly available statistical information. Finally, we perform several case studies on the state of urban resource allocations such as public transport (Divvy bikes, CTA trains, and buses), schools, and hospitals in the city of Chicago using the \system-generated data.
{Our experiments revealed a greater degree of equity in the allocation of schools, hospitals, and CTA bus stations, compared to the allocation of Divvy bikes and the CTA L train, which exhibited a high level of disparity.}
Furthermore, perhaps contrary to expectations, our results show a significant {disadvantage for} the Whites group.
In addition to equity evaluation at the aggregate level, \system can be used for providing {\em visual explanations} for observed inequities. In particular, we use a set of samples generated by \system and generate a map that reveals the Divvy bikes inequities are due to a significantly lower access levels to the stations in the west and northwest neighborhoods of Chicago. 




\noindent{\bf Summary of contributions.} In summary, our contributions
in this paper are as follows:
\begin{itemize}[leftmargin=*]
    \item We propose \system, a system for generating simulated individual-level population data that benefits from the publicly available aggregated demographic statistics.
    
    \item \system enables a wide range of applications requiring individual-level population data with demographic information. Two specific applications of \system include 1) creating semi-synthetic datasets that can be used for a variety of tasks, such as auditing equitable resource allocation, 2) Enabling the development of randomized algorithms for social applications for the urban population.
    
    \item We generated semi-synthetic benchmark datasets for Chicago and validated them through statistical tests{, which serve as a tool to evaluate the equity of the allocation of city resources}. 

    \item We perform several interesting case studies investigating the equitable allocation of urban resources, such as {\it CTA trains, buses, Divvy bikes, schools}, and {\it hospitals} among different demographic groups. 
\end{itemize}

\section{System Overview}\label{sec:sys}
We aim to develop a system for simulating individual-level population data from publicly available city statistics that can be used for auditing the equitable allocation of resources.
In particular, we would like to enable a variety of features that may be used to generate both semi-synthetic benchmarking datasets and individual sample generation for sampling-based approaches.

\begin{figure}[!tb]
    \centering
    \includegraphics[width=.47\textwidth]{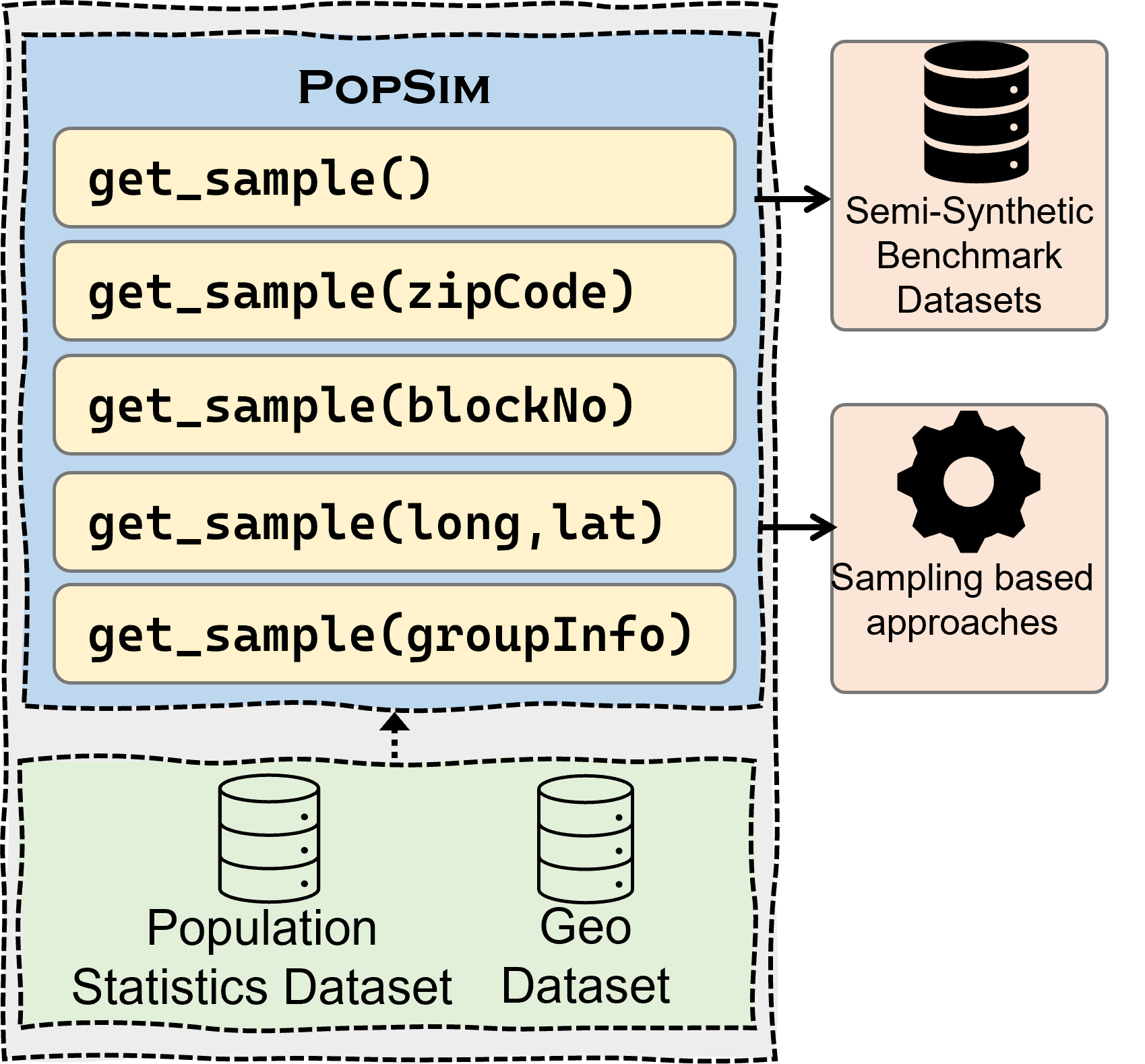}
    \caption{System Overview}
    \label{fig:sys}
\end{figure}

Figure~\ref{fig:sys} shows the high-level overview of our system.
\system uses two types of publicly available data as input in its core: (i) {\em \stat dataset} (\S~\ref{sec:data:stat}) that provides the statistical information,
and (ii) {\em \geo databases} (\S~\ref{sec:data:geo}) that is used for identifying the boundaries of geo-regions such as a block and zip code.
\system provides a set of functionalities to sample from different granular levels of geo-location hierarchies and demographic groups. Each sample is a tuple in form of {\bf $[\langle$ {\tt\small long, lat}$\rangle,\langle$ {\tt\small groupInfo} $\rangle]$}, containing the  sample location and its demographic information ({\tt\small race}, {\tt\small gender}, etc.).
\begin{itemize}[leftmargin=*]
    \item {\tt\small get\_sample()}: returns an unbiased sample based on the overall population distributions provided by \stat dataset.
    \item {\tt\small get\_sample(zipCode/blockNo)}: returns an unbiased sample from the specified zip code or block\footnote{While being a standard notion as a fine-grained aggregate level in US cities, different countries may have other geographical units.
    Without loss of generality, in this paper, we used the US addresses system for explaining \system. To adapt \system for countries with different standards, it is enough to replace ``block'' with the most fine-grained geographical unit for which statistical aggregates are publicly available.
    }.
    \item {\tt\small get\_sample(long,lat)}: returns an unbiased sample with the location as specified by the input longitude and latitude.
    \item {\tt\small get\_sample(groupInfo)}: returns an unbiased sample from the specified demographic group (e.g. {\tt\small race}), or a set of demographic groups, based on the overall distribution of that group.
\end{itemize}

\system enables a wide range of applications that require individual-level population data with demographic information.
Two specific applications of \system are:
\begin{itemize}[leftmargin=*]
    \item {\em Semi-synthetic datasets:} Repeated sampling from \system provides benchmark datasets with individual-level demographic information that can be used for different tasks, such as auditing equitable allocation of city resources. For example, if one would like a dataset with $n$ samples from the entire city (resp. a demographic group), it is enough to generate $n$ samples using the {\tt\small get\_sample()} (resp. {\tt\small get\_sample(groupInfo)}).
    To demonstrate this, in this paper, as we shall further explain in \S~\ref{sec:benchmark}, we generated several datasets with various sizes from the city of Chicago and used them in \S~\ref{sec:casestudy} for auditing equity in the allocation of various resources across the city.
    \item {\em Sampling-based approaches:} Unbiased sampling is a key requirement for Randomized Algorithms~\cite{motwani1995randomized} and Monte-Carlo methods~\cite{hammersley2013monte}. \system empowers developing sampling-based approaches for social applications for the city population. 
    An example of such applications is fair allocation of resources, studied by~\cite{blanco2022fairness,asudeh2022maximizing} (see \S~\ref{sec:related}).
\end{itemize}

\vspace{1mm}\noindent {\bf Implementation Details and Artifact availability.}
\system is an open source system, implemented {\tt\small Python}, using {\tt\small pandas}, {\tt\small numpy}, and {\tt\small geopandas} packages. In addition, we use {\tt\small seaborns} and {\tt\small matplotlib} libraries for the analytical step.
The code is publicly available on GitHub.\footnote{\url{https://github.com/UIC-InDeXLab/PopSim}}
As we shall explain in \S~\ref{sec:benchmark}, we use \system to generate several semi-synthetic datasets for the city of Chicago. The datasets are also publicly available\footnotemark[1].
\section{Preliminaries}\label{sec:pre}
\subsection{Sample Generation Techniques}\label{sec:pre:sample}
We mainly use two sample generation techniques for simulating the individual-level data from the publicly available data: 
\begin{itemize}[leftmargin=*]
    \item {\em Inverse-CDF Sampling} \cite{devroye1986sample}.
        Also known as Inverse transform sampling, inverse-CDF is an approach for generating random samples from a given probability distribution. 
        Let $f(x)$ be the reference probability density function (PDF). To draw samples from $f$, the inverse-CDF approach first computes the cumulative density function (CDF) of $f$, as $F(x)=P(X \le x)=\int_{-\infty}^x f(x) dx$. 
        It then computes the inverse CDF function $F^{-1}(x)$ and uses it for generating the samples.
        We will further explain this approach in \S~\ref{sec:methods:1} for sampling from large regions.
        Indeed this approach is limited to the distributions for which the inverse CDF is commutable apriori.
    \item {\em Monte Carlo Rejection Sampling} \cite{raeside1976monte}.
        This approach is useful for generating unbiased samples from a probability distribution with an odd-shaped probability density function $\psi$ that is challenging to sample directly. 
        The core idea behind this technique is to first, find a tight and ``simple-to-sample'' distribution (usually the bounding box) $\xi$ that encloses $\psi$. Then, instead of sampling from $\psi$, it generates a sample from $\xi$. The sample is accepted if it falls under the curve of $\psi$. Otherwise, the sample is rejected (no sample is generated) and the process repeats. 
        We use Monte Carlo rejection sampling in \S~\ref{sec:methods:2} for generating unbiased samples within a block.
\end{itemize}
\textcolor{blue}{
}

\subsection{Input Datasets}\label{sec:inputData}

{\system takes two types of datasets as the input:}
\begin{itemize}
    \item {\it Population statistics datasets:} These datasets provide information about the population distribution across the city. Some of the well-known examples of such data include Decennial Census, American Community Survey (ACS), American Housing Survey (AHS), Population Projections, etc.
    \item {\it Geo databases:} These datasets provide spatial data such as geographic boundaries, with different levels of granularity, used to identify geo-regions varying from block to national level coverage.
\end{itemize}

Without loss of generality, in this paper, we fine-tune our system for the city of Chicago based on the following publicly available datasets. 

\subsubsection{Decennial Census of Population and Housing Data Database 2020}\label{sec:data:stat}
Decennial Census is a census of the population of the country that is conducted every 10 years, ending in a zero~\cite{bureau_2021}. The census counts each resident of the country, recording where they reside on April 1st.
The data contains fine-grained information at the {\bf block level}, containing 
residency information such as the size of the household and the type of residence,  
 as well as demographic characteristics of the population, including {\tt\small gender}, {\tt\small race and ethnicity}, and {\tt\small age}.
Census results are widely used for tasks such as determining the distribution of seats in the House of Representatives, shaping the boundaries of congressional districts, and annual allocation of federal funding.
In this paper, we use the decennial data for the state of Illinois and more specifically the city of Chicago. We primarily focus on race-based statistics.

\subsubsection{TIGER/Line Geodatabases 2022}\label{sec:data:geo}
The most detailed geospatial data for mapping census demographic statistics are the TIGER/Line data from the U.S. Census Bureau's Topologically Integrated Geographic Encoding and Referencing System~\cite{tiger_2022}.
In this paper, we specifically, use the TIGER/Line geodatabase for identifying the boundaries of the blocks across the state of Illinois. While each block has several properties, we only use the Federal Information Processing Series (FIPS) code and the border geometry of Illinois blocks. FIPS codes 
are assigned to various geographic entities such as states, counties, metropolitan areas, cities, county subdivisions, consolidated cities, and indigenous areas, based on their alphabetical names.


\section{System Development Details}\label{sec:methods}
After providing a high-level overview of our system and the preliminaries, in this section, we provide the development details of \system.
In particular, in \S~\ref{sec:methods:1} 
 we first discuss sampling from the entire population, a specific zip code, or a demographic group.
Next, in \S~\ref{sec:methods:2}, we provide the details for sampling from the finest granularity levels, i.e., from a specific block or coordinate.

\subsection{Sampling from A Large Region}\label{sec:methods:1}
Inverse-CDF is the core idea for sampling from regions larger than a block, i.e., a specific zip code or the entire population.
Recall from \S~\ref{sec:data:stat} that the \stat dataset contains fine-grained statistics at the {\em block} level.
Therefore, in order to sample from a specific region $R$ (e.g., a zip code), Algorithm~\ref{alg:1} first samples a block within the specific region, with the probability density proportional to its size and then returns a sample from the selected block.

\begin{algorithm}[!htb]
\caption{}\label{alg:1}
\begin{algorithmic}[1]
\Function{get\_sample}{zipCode=null}
    \Statex {\tt\small // find the set of blocks in given zipCode}
    \State $B\gets$ {\sc get\_blocks}(zipCode)
    \Statex {\tt\small // compute the cumulative function}
    \State $F[0]\gets 0$; $sum\gets 0$
    \For{$i\gets 1$ to $|B|$}
        \State $F[i]\gets F[i-1]+B[i].\mbox{population}$
        \State $sum\gets sum+B[i].\mbox{population}$
    \EndFor
    \For{$i\gets 1$ to $|B|$}
        $F[i]\gets F[i]/sum$
    \EndFor
    \Statex {\tt\small // find the block to sample next}
    \State $u\gets U[0,1]$ {\tt\small // random uniform in range [0,1]}
    \State block $\gets$ {\sc binary\_search}(F,u)
    \State {\bf return} {\sc get\_sample}(block) {\tt\small //Algorithm~\ref{alg:2}}
\EndFunction
\end{algorithmic}
\end{algorithm}

To further clarify how Algorithm~\ref{alg:1} works, let us consider a toy example, where the selected region contains the following blocks with the specified populations:
\begin{center}
    \begin{tabular}{|c|c|c|c|c|c|c|}
    \hline
         blockNo& 1 & 2 & 3 & 4 & 5 & 6 \\ \hline
         population& 87 & 230 & 310 & 112 & 167 & 94  \\ \hline
    \end{tabular}
\end{center}
Following the lines 3 to 7 of Algorithm~\ref{alg:1}, the vector of the cumulative function $F$ is computed as
\begin{center}
    \begin{tabular}{|c|c|c|c|c|c|c|}
    \hline
         blockNo& 1 & 2 & 3 & 4 & 5 & 6 \\ \hline
         $F$& .087 & .317 & .627 & .739 & .906 & 1  \\ \hline
    \end{tabular}
\end{center}
Next, the algorithm draws a random uniform number in the range $[0,1]$. Suppose the generated random number is $0.786$.
Since 0.786 is larger than 0.739 and smaller than 0.906, the binary search on $F$ with $0.786$ returns blockNo 5.
Finally, the algorithm calls Algorithm~\ref{alg:2} to draw an unbiased sample from block 5.

To draw a sample from a specific demographic group or a set of groups, {\tt\small groupInfo}, one needs to first update the block populations to only include the counts for {\tt\small groupInfo}. It should then limit the demographic groups of the selected block to {\tt\small groupInfo} before calling Algorithm~\ref{alg:2} to sample it.

\subsection{Sampling from a specific block or location}\label{sec:methods:2}
Drawing an unbiased sample from a specific block requires (a) identifying the demographic information of the selected sample and (b) assigning a specific location ({\tt \small long, lat}) to it.

\begin{algorithm}[!htb]
\caption{}\label{alg:2}
\begin{algorithmic}[1]
\Function{get\_sample}{blockNo}
    \Statex {\tt\small // (a) Find the groupNo of the selected sample}
    \State g $\gets$ {\sc get\_group}(blockNo) {\tt\small //Algorithm~\ref{alg:3}}
    \Statex {\tt\small // (b) specify the sample location}
    \State $(x_\triangleleft, x_\triangleright ) \gets$ the minimum \& maximum longitude of block[blockNo]
    \State $(y_\bigtriangledown,y_\bigtriangleup ) \gets$ the minimum \& maximum latitude of block[blockNo]
    \State reject$\gets${\bf true}
    \While{reject}
        \State long $\gets x_\triangleleft + U[0,1](x_\triangleright - x_\triangleleft)$
        \State lat $\gets y_\bigtriangledown + U[0,1](y_\bigtriangleup - y_\bigtriangledown)$
        \If{{\sc IsInside}((long,lat), block[blockNo])}
            \State reject$\gets${\bf false}
        \EndIf
    \EndWhile
    \State {\bf return} $([long,lat],g)$
\EndFunction
\end{algorithmic}
\end{algorithm}

In Algorithm~\ref{alg:2}, we use Inverse-CDF for (a), similar to our approach in Algorithm~\ref{alg:1}.
However, since the block boundaries do not form standard geometric shapes, we devise Monte-Carlo rejection sampling (\S~\ref{sec:pre:sample}) for (b).
To do so, the algorithm first creates the tight bounding box around the specified block (lines 3 and 4). It then generates uniform random samples within the box and accepts them (line 10) if it falls inside the block.

In order to generate a sample from a specific location ({\tt\small get\_sample} in Algorithm~\ref{alg:3}), we first need to identify the corresponding block for the location. Then, it is enough to call {\tt\small get\_group} function to sample the demographic information of the selected sample. 

\begin{algorithm}[!htb]
\caption{}\label{alg:3}
\begin{algorithmic}[1]
 \Function{get\_group}{blockNo}
    \State groups$\gets$ block[blockNo].groups
    \State $F[0]\gets 0$; $sum\gets 0$
    \For{$i\gets 1$ to $|\mbox{groups}|$}
        \State $F[i]\gets F[i-1]+\mbox{groups}[i].\mbox{population}$
        \State $sum\gets sum+\mbox{groups}[i].\mbox{population}$
    \EndFor
    \For{$i\gets 1$ to $|\mbox{groups}|$}
        $F[i]\gets F[i]/sum$
    \EndFor
    \State $u\gets U[0,1]$ {\tt\small // random uniform in range [0,1]}
    \State {\bf return} {\sc binary\_search}(F,u)
\EndFunction
\Statex
 \Function{get\_sample}{long, lat}
    \State blockNo$\gets$ {\sc block(long,lat)}
    \State {\bf return} $([long,lat],{\sc get\_group}(blockNo))$
\EndFunction
\end{algorithmic}
\end{algorithm}


\subsection{System Extension}
While we demonstrate \system using the publicly available data sets for Chicago, its scope is indeed not limited to this city.
First, to fine-tune \system for a different city in the US, it is enough to use the population statistics (Census Population Data) and geo-boundary databases of the target city.
Tuning \system for other countries with different geographical units and aggregations statistics require replacing the notions such as zip-code and block to the standard notions in the target country.

It is easy to use heterogeneous statistical data sets to augment samples generated by \system with additional attributes.
As an example, suppose one would like to add two columns {\tt\small income} and {\tt \small job title} to each sample. Note that  {\tt\small income} is continuous ordinal while {\tt \small job title} is non-ordinal categorical.
Data sets that provided this information at some aggregate level are publicly available (e.g., \cite{chicagoDemog}).
The first step to augment a sample with additional data is to identify which unit it belongs to. For example, suppose the {\tt\small income} and {\tt \small job title} data are provided at the block level. Then given a sample, we should first identify which block it belongs to.
Next, we should sample the attribute values according to the distribution of the given unit.
For non-ordinal categorical attributes such as {\tt \small job title}, one can use Inverse-CDF (similar to Algorithm~\ref{alg:1} to draw an unbiased value (e.g., {\tt \small job title: educator}). For ordinal continuous attributes, on the other hand, one can use Normal distribution (with the average and variance specified for the given unit) for specifying the attribute value (e.g., {\tt\small income: \$98,450}).


\section{Benchmark Datasets}\label{sec:benchmark}
As previously mentioned in \S~\ref{sec:sys}, one of the applications of \system is to build semi-synthetic benchmark datasets.
To demonstrate this, we generated six population datasets for the state of Illinois with \system. The largest dataset has an identical population size to that of the state of Illinois in the Decennial Census of Population dataset, containing $n=12,812,508$ samples. The other datasets are of size $5M$, $1M$, $500K$, $200K$, and $50K$ samples, respectively. 
We shall later use these datasets for our case studies in \S~\ref{sec:casestudy}. But first, we use statistical tests in \S~\ref{sec:benchmark:validation} to confirm that the generated datasets indeed follow the underlying distribution by the input \stat data.


\subsection{Datasets Validation using Statistical Tests}\label{sec:benchmark:validation}
In order to validate the generated datasets, we must ensure that they exhibit an identical distribution to that of the input \stat dataset (Decennial Census of Population). Therefore,
we establish the {\em null hypothesis}  of ``there is no substantial difference in population distribution between any of the \system-generated datasets and the Decennial Census of Population dataset'', and proceed with a few statistical tests to reject it.

We start by normalizing both synthetic datasets and the Decennial Census of Population dataset, ensuring that our assessments concentrate on the underlying distributions of the data rather than the size of the datasets. Next, we compute the mean ($\mu$) and standard deviation ($\sigma$) of the datasets, enabling a comparison of the comparability of central tendency and dispersion across the various datasets.

We compare the distributions of each generated dataset with the Decennial Census of Population dataset based on two different properties. 
\begin{itemize}[leftmargin=*]
    \item {\em Demographic Information:} We compare the distributions of each demographic group in the synthetic datasets with the total population of that group within the population statistics dataset.
    \item {\em Block:} We compare the population of each block in the synthetic datasets with the geo-distribution of that block according to the Decennial Census of Population dataset. Due to the significant variation in the block FIPS column, we combined every $m$ adjacent block into one prior to performing the statistical tests. The combined number of blocks $m$ varies inversely with the size of the synthetic dataset. Specifically, for datasets of size 12.8M, 5M, 1M, 500K, 200K, and 50K, $m$ takes on the values of 5, 10, 50, 100, 250, and 1000, respectively.
\end{itemize}
We used the {\bf Kolmogorov–Smirnov (K-S) test} and {\bf t-test} to compare each synthetic dataset with the Decennial Census of Population dataset and test the null hypothesis. The results of the population comparison w.r.t {\em demographic group} (race) and {\em block} properties are presented in Tables \ref{tab:first} and \ref{tab:second} and confirm that the means values are not substantially different from those of the Decennial Census of Population dataset as evidenced by most of the p-values falling over the 0.05 threshold.
 
Table \ref{tab:second} shows a few exceptions when comparing based on the block property. For instance, according to the K-S test, the first three synthetic datasets have significantly different distributions from the Decennial Census of Population dataset, as their respective p-values are less than 0.05. However, the p-values of the three larger datasets suggest otherwise, indicating that they can be used with confidence for any related tasks.

For the race attribute, as demonstrated in Table \ref{tab:first}, the K-S test results imply that with the exception of the first synthetic dataset (size 50K), the distributions of the remaining datasets are comparable to the Decennial Census of Population dataset.

Overall, the results indicate that, with the exception of the first three synthetic datasets, the null hypothesis can be rejected with high confidence. Therefore, the generated datasets are viable synthetic options for any task in need of individual data with demographic and geopositioning information.



\begin{table}[]
\centering
\resizebox{\columnwidth}{!}{%
\begin{tabular}{ccccc}
\hline
\multirow{2}{*}{Datasets}                                                     & \multicolumn{2}{c}{t-test} & \multicolumn{2}{c}{K-S Test} \\ \cline{2-5} 
                                                                              & Statistic     & p-value    & Statistic      & p-value     \\ \hline
50,000                                                            & 0.50288       & 0.61611    & 0.40625        & 0.000312     \\
200,000                                                            & 0.21103       & 0.83323    & 0.21698        & 0.10847     \\
500,000                                                            & 0.09070       & 0.92787    & 0.11017        & 0.79395     \\
1,000,000                                                          & 0.05354       & 0.95738    & 0.10656        & 0.81513     \\
5,000,000                                                          & 0.0           & 1.0        & 0.06250        & 0.99972     \\
12,854,526                                                         & 0.01756       & 0.98601    & 0.05556        & 0.99971     \\
\hline
\end{tabular}%
.,/.}
\caption{Statistical tests by race}\label{tab:first}

\centering
\resizebox{\columnwidth}{!}{%
\begin{tabular}{ccccc}
\hline
\multirow{2}{*}{Datasets}                                                     & \multicolumn{2}{c}{t-test} & \multicolumn{2}{c}{K-S Test} \\ \cline{2-5} 
                                                                              & Statistic    & p-value     & Statistic      & p-value     \\ \hline
50,000                                                            & 2.44894      & 0.999998   & 0.20317          & 4.21e-179     \\
200,000                                                            & -6.7305      & 0.9999946   & 0.07142        & 2.10e-22     \\
500,000                                                            & -4.9672      & 0.9999960   & 0.02846        & 0.00037     \\
1,000,000                                                          & -3.4843      & 0.9999972   & 0.01115        & 0.39499     \\
5,000,000                                                          & -3.1607      & 0.9999974   & 0.00681        & 0.65584     \\
12,854,526                                                         & -3.5042      & 0.9999972   & 2.10259        & 1.0000      \\
\hline
\end{tabular}%
}
\caption{Statistical tests by block (FIPS)}\label{tab:second}
\end{table}

\section{Case Study}\label{sec:casestudy}
Having verified the validity of the synthetic datasets generated by \system, in this section, we perform several interesting case studies on the state of urban resource allocations in the city of Chicago. Specifically, we investigate the {\em equitable allocation} of the following urban resources across the city for difference {\em racial groups}: {\tt\small (1) hospitals\footnote{https://data.cityofchicago.org/Health-Human-Services/Hospitals-Chicago/ucpz-2r55 }, (2) schools\footnote{https://data.cityofchicago.org/Education/Chicago-Public-Schools-School-Locations-SY2021/p83k-txqt/data}, (3) Divvy bikes stations\footnote{https://data.cityofchicago.org/Transportation/Divvy-Bicycle-Stations-In-Service/67g3-8ig8}, (4) CTA train stations\footnote{https://data.cityofchicago.org/dataset/CTA-L-Rail-Stations-kml/4qtv-9w43 },} and {\tt\small (5) bus stops\footnote{https://data.cityofchicago.org/Transportation/CTA-Bus-Stops-kml/84eu-buny }}. Due to space limitations, we only focus on our largest dataset (with a population of 12,854,526 samples). 

We define the ``accessibility'' of a resource as the {\em euclidean distance}\footnote{We transform the EPSG:4326 geographic coordinate system into the EPSG:26916 local projected coordinate system prior to calculating the metrics. This allows us to utilize the Euclidean distance formula to compute the distance with a 2.0 meters error in the region of Illinois \cite{epsg}. We admit that actual route distance (walking distance) is more precise for computing the distance to resources. However, for simplicity, we use Euclidean distance.} from an individual's geolocation to the closest resource of that particular type.
We use the spatial KD-tree indexing~\cite{ooi1987spatial} for locating the closest resource to each person. 
In order to evaluate racial equity in resource allocation, we compare the average distance-to-closest-resource  for different groups. 
Formally, for each group $\mathbf{g}$ and the resource locations $R$, the average distance is computed using Equation~\ref{eq:avgdist}.
\begin{align}\label{eq:avgdist}
    \delta_R(\mathbf{g}) = \frac{1}{|\mathbf{g}|}\sum_{t\in \mathbf{g}} \min_{r\in R} \big(\mbox{dist}(t,r)\big)
\end{align}

\begin{figure*}[!ht] 
\begin{minipage}[t]{\linewidth}
\centering
\begin{subfigure}[t]{0.32\linewidth}
\centering
  \includegraphics[width=\textwidth]{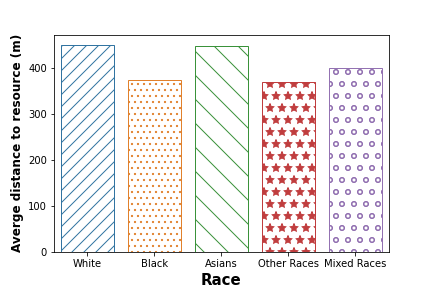}
  \caption{Schools}
  \label{fig:graphfig1}
\end{subfigure}
\hfill
\begin{subfigure}[t]{0.32\linewidth}
\centering
  \includegraphics[width=\textwidth]{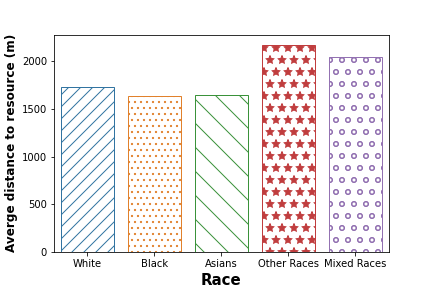}
  \caption{Hospitals}
  \label{fig:graphfig2}
\end{subfigure}
 \hfill
\begin{subfigure}[t]{0.32\linewidth}
\centering
  \includegraphics[width=\textwidth]{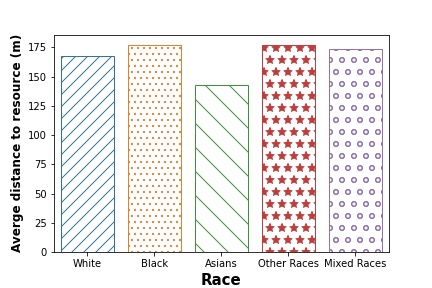}
  \caption{Bus stops}
  \label{fig:graphfig3}
\end{subfigure}

\hspace{5mm}
\begin{subfigure}[t]{0.32\linewidth}
\centering
  \includegraphics[width=\textwidth]{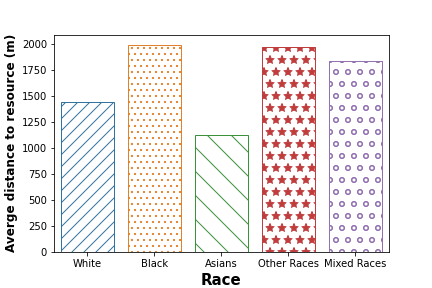}
  \caption{CTA train stations}
  \label{fig:graphfig4}
\end{subfigure}
\hspace{5mm}
\begin{subfigure}[t]{0.32\linewidth}
\centering
  \includegraphics[width=\textwidth]{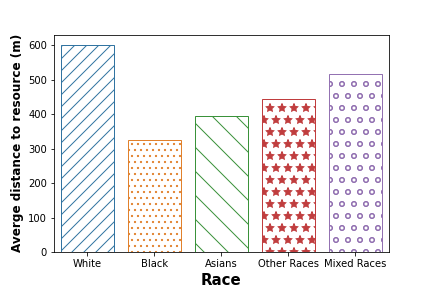}
  \caption{Divvy bike stations}
  \label{fig:graphfig5}
\end{subfigure}
\end{minipage}
\vspace{-2mm}
\caption{The Average distance-to-closest-resource for different racial groups and various urban resources in the city of Chicago. For presentation purposes, the populations of other races are combined as {\it ``Other Races''}. {\it ``Mixed Races''} includes all multi-racial population.} \label{fig:graphfigs}
\end{figure*}

\subsection{Experiment Results}

Figure~\ref{fig:graphfigs} shows our experiment results on
the average distance-to-closest-resource (in meters) for different racial groups and urban resources in Chicago.
Looking at the figure one can observe inequities in resource accessibility across all cases.
On average, Asians had the highest distance to the closest school (449.50m), although they traveled the lowest distances to the nearest bus stop (142.90m) and CTA train station (1121.3m). Conversely, Black individuals walked the highest distance to the nearest bus stop (177.08m) and CTA train station (1992.6m), but the least distance to the nearest Divvy bike station (324.31m) and hospital (1636.35m). The maximum average distance traveled to the Divvy bike station was by Whites (601.44m), while the maximum for the nearest hospital was by Other Races (2170.20m).

We measure the inequities in the form of the {\em maximum disparity ratio}. For each resource $R$ (e.g., bus stops) and the groups $G$, the maximum disparity ratio is computed using Equation~\ref{eq:disp} and reported in Table~\ref{tab:disp}.
\begin{align}\label{eq:disp}
    \mbox{disparity}(R)= \frac{\max_{\mathbf{g}\in G } (\delta_R(\mathbf{g}))}{ \min_{\mathbf{g'}\in G } (\delta_R(\mathbf{g'}))}
\end{align}

\begin{table}[!hbt]
    \centering
    \begin{small}
    \begin{tabular}{@{}c@{}|@{}c|@{}c|@{}c|@{}c|c@{}}
         &{\tt\small schools}&{\tt\small hospitals}&{\tt\small bus}&{\tt\small train}&{\tt\small Divvy}  \\\hline
         {\small min}& 370.04 & 1636.35 & 142.90 & 1121.32 &  324.31 \\
         {\small max}& 449.50 & 2170.20 & 177.08 & 1992.61 & 601.44  \\ \hline 
         {\small disparity}& 1.215 & 1.326 & 1.239 & 1.777 & 1.855 \\
    \end{tabular}
    \end{small}
    \caption{Resource allocation disparities}
    \label{tab:disp}
\end{table}

Among the evaluated resources, allocation in {\tt\small schools}, {\tt\small hospitals}, and {\tt\small bus stops} (Figures~\ref{fig:graphfig1}, ~\ref{fig:graphfig2}, and ~\ref{fig:graphfig3}) were more equitable as 
the maximum disparity ratio
was smaller.
On the other hand, allocation of {\tt\small CTA Train stations} and {\tt\small Divvy bike stations} (Figures~\ref{fig:graphfig4} and~\ref{fig:graphfig5}) were more inequitable. In particular, {\em Divvy bike stations} had the maximum inequity, where the average distance of the White individuals to the closest station is 85\% higher than that of Black individuals.

\begin{figure}
\centering
\includegraphics[width=.9\linewidth]{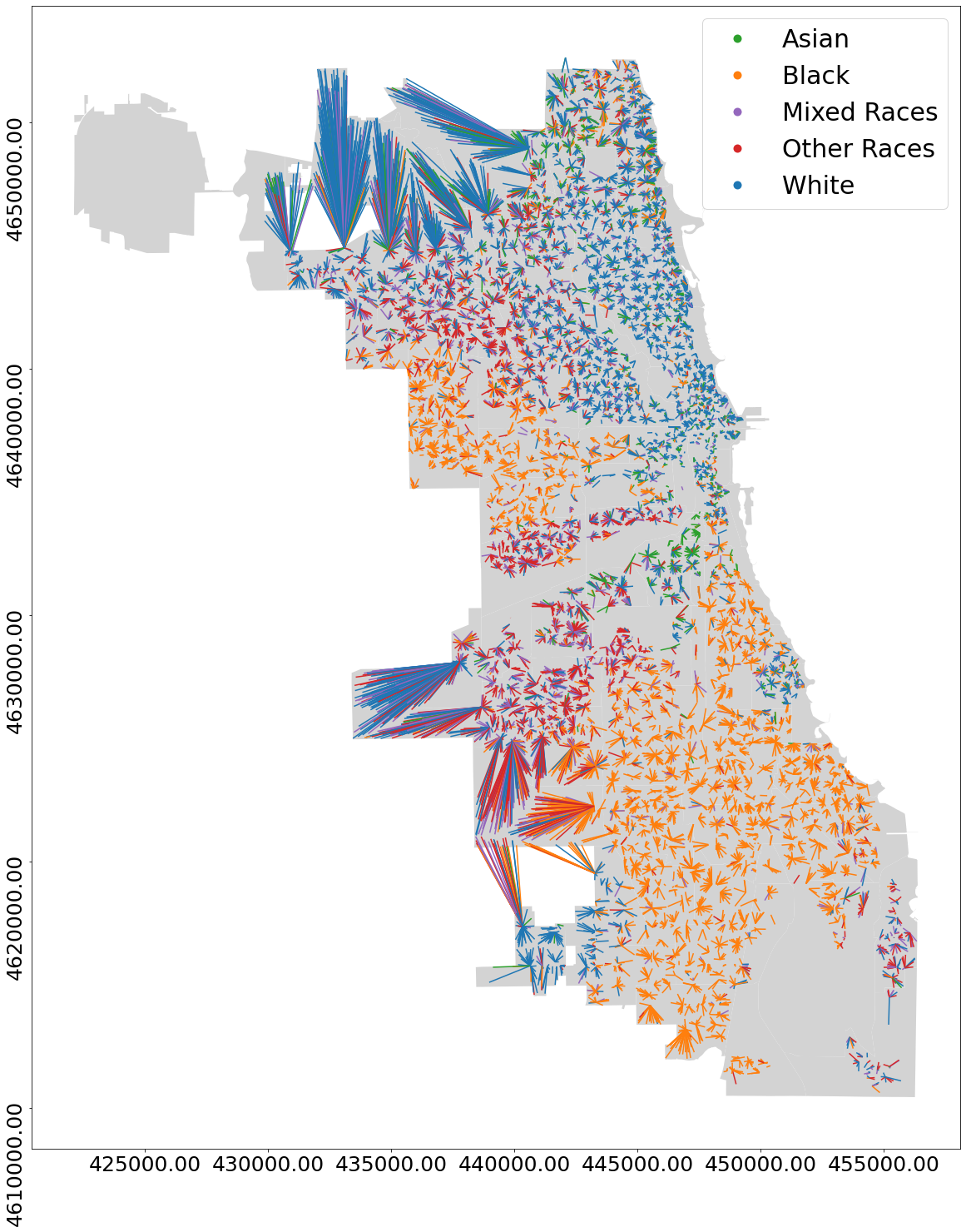}
\caption{Visual Explanation of inequity in Divvy bike stations accessibility.}\label{fig:exp:visual}
\end{figure}

In addition to evaluating inequity, our datasets can be used for providing {\bf visual explanations} for the inequities.
In our experiments, we observed the maximum inequity for Divvy bike stations. Besides, perhaps counter-intuitively, we observed inequity against White individuals.
Therefore, to demonstrate using our datasets for visual explanations, we generated Figure~\ref{fig:exp:visual} by generating 2000 random samples from Chicago using \system and connecting each sample to its closest Divvy Bike Station on the map.
From the figure, it is evident that some of the mostly-White neighborhoods in the northwest (and west) of Chicago (with long blue lines) have caused the inequity. It turns out there is no Divvy bike station near those regions which caused an increase in the average travel distances for the White group.

Further investigating this issue, we realized that it happened due to the current phase of the Divvy expansion plan. The first phase of expansion focused on underrepresented communities in South Chicago, while the second phase, which began in 2021 and continues to the present, targets North West and South West communities\cite{DivvyBikes}. The third phase, scheduled for the near future, will focus on the Northwest and Southwest areas. Furthermore, there has been significant interest expressed by users requesting new bike stations, as evidenced by a large number of requests on the proposed Divvy bike station map\cite{DivvyLocationsChicago}. In particular, there is considerable demand for Divvy stations in North West communities like Jefferson Park, where there are currently no Divvy stations.
\section{Related Works}\label{sec:related}
Resource allocation problems have been extensively studied for decades in various disciplines namely, economics, management, healthcare, urban planning, computer science, etc. With the recent emergence of topics on fairness in computational problems, socially fair, just, and equitable resource allocation has drawn lots of attention. 
In \cite{blanco2022fairness}, authors study the problem of fair resource allocation in the context of location problems where they try to determine the position of one or more facilities to satisfy the demand of a set of users while satisfying the fairness from the facilities' perspective. Similarly, in \cite{asudeh2022maximizing}, they analyze a covering location problem with fairness
constraints minimizing the pairwise deviations between the different covered sets. In \cite{lan2010axiomatic}, authors propose a set of 5 axioms for fairness measures based on which they construct a family of fairness measures for network resource allocation.

Many fairness-aware solutions for computational problems have been proposed in the past decade. These solutions usually fall into one of the categories of fairness-related interventions in the data, modifying the training process of the learning algorithms, and altering the outcomes of the models. Regardless of the level, at which the interventions are applied, they need to be evaluated empirically on benchmark datasets that represent realistic and diverse settings \cite{le2022survey}.

\section{Final Remarks}\label{sec:conclusion}


In this paper, we introduced \system and used it to generate effective high-proximity benchmark datasets. The result datasets may be utilized for population statistical research and serve as benchmarking datasets for recent fairness-aware solutions.
As a disclaimer, we believe the inequalities found by our system and in our case study (\S~\ref{sec:casestudy})
should not necessarily be seen as an indication of inequitable resource distributions, as other factors and criteria may also have been considered for resource allocation.
For instance, different groups may have different demands on various resources and may prefer some resource types over others.
In such settings, an equitable allocation of resources that equally satisfies the demands of different groups may be different from equal access to all resources.


\bibliographystyle{siam}
\bibliography{ref}
\balance

\end{document}